%% file: draft.tex
\documentclass[cernpreprint,texlive=2011,txfonts,UKenglish]{latex/atlasdoc}
\pdfoutput=1

\usepackage[lineno=false,biblatex=true]{latex/atlaspackage}
\usepackage{latex/atlasbiblatex}

\usepackage{latex/atlascontribute}

\usepackage{latex/atlasphysics}

\graphicspath{{logos/}{figures/}}

\usepackage{draft-defs}

\usepackage{multirow}

\usepackage{morefloats}

\input{draft-metadata}

\hypersetup{pdftitle={Search for new phenomena in dijet mass and angular distributions with the ATLAS detector at sqrt(s)=13 TeV},pdfauthor={The ATLAS Collaboration}}

\begin{document}

\maketitle

\pagebreak

\input{main}

\printbibliography

\newpage 
\input{atlas_authlist}

\end{document}

%% file: draft-metadata.tex
\AtlasTitle{Search for new phenomena in dijet mass and angular distributions from $pp$ collisions at $\sqrt{s}$ = 13 \TeV\xspace with the ATLAS detector}

\author{The ATLAS Collaboration}

\date{\today}

\PreprintIdNumber{CERN-PH-EP-2015-311}

\AtlasDate{\today}

\AtlasJournal{Phys. Lett. B}

\AtlasAbstract{
This \publicationType describes a model-agnostic search for pairs of jets (dijets) produced by resonant and non-resonant phenomena beyond the Standard Model in \integLumi of proton--proton collisions with a centre-of-mass energy of $\sqrt{s}=13$~TeV recorded by the ATLAS detector at the Large Hadron Collider. The distribution of the invariant mass of the two leading jets is examined for local excesses above a data-derived estimate of the smoothly falling prediction of the Standard Model. The data are also compared to a Monte Carlo simulation of Standard Model angular distributions derived from the rapidity of the two jets.
No evidence of anomalous phenomena is observed in the data, which are used to exclude, at 95\% CL, quantum black holes with threshold masses below \qbhLimitMjj, \blackMaxLimitMjj, or \RSLimitMjjExpected in three different benchmark scenarios; resonance masses below \qstarLimitMjj for excited quarks, \wprimeLimitMjjLow in a $W^\prime$ model, \zprimeMassCouplingMjj\xspace in a $Z^\prime$ model; and contact interactions with a compositeness scale below \LambdaLimitPlusChi and \LambdaLimitMinusChi respectively for destructive and constructive interference between the new interaction and QCD processes. These results significantly extend the ATLAS limits obtained from 8 TeV data. 
Gaussian-shaped contributions to the mass distribution are also excluded if the effective cross-section exceeds values ranging \gaussianRangeStatement. 
}

%% file: main.tex
\section{Introduction}
\label{sec:intro}

\input{01intro.tex}

\section{The ATLAS detector}
\label{sec:detector}

\input{02detector.tex}

\section{Data selection}
\label{sec:selection}

\input{03selection.tex}

\section{Simulated collisions}
\label{sec:simulation}

\input{04simulation.tex}

\section{Selection for the mass distribution analysis}
\label{sec:resonance}

\input{05resonance.tex}

\section{Selection for the angular distributions analysis}
\label{sec:angular}

\input{06angular.tex}

\section{Signal models}
\label{sec:bsm}

\input{07signals.tex}

\section{Limits}
\label{sec:limits}

\input{09limits.tex}

\section{Conclusion}
\label{sec:conclusion}

\input{10conclusions.tex}

\section*{Acknowledgements}
\label{sec:acknowldge}

\input{acknowledgements/Acknowledgements.tex}

\clearpage

%% file: 01intro.tex
The centre-of-mass energy of proton--proton ($pp$) collisions at the Large Hadron Collider (LHC) at CERN has been increased from $\sqrt{s}=8$~\TeV\xspace to $\sqrt{s}=13$~\TeV\xspace, opening a new energy regime to observation.

New particles produced in LHC collisions must interact 
with the constituent partons of the proton.
Consequently, the new particles can also produce partons in the final state.
Final states including partons often dominate in models of new phenomena beyond the Standard Model (BSM).
The partons shower and hadronize, creating collimated jets of particles carrying approximately the four-momenta of the partons.
The total production rates for two-jet (dijet) BSM signals can be large, allowing searches for anomalous dijet production to test for such signals with a relatively small data sample, even at masses that constitute significant fractions of the total hadron collision energy.

In the Standard Model (SM), hadron collisions produce jet pairs primarily via $2\rightarrow 2$ parton scattering processes governed by quantum chromodynamics (QCD).
Far above the confinement scale of QCD ($\approx$ 1~\GeV\xspace), jets emerge from collisions with large transverse momenta, \pT, perpendicular to the direction of the incident partons.
For the data analysed here, QCD predicts a smoothly falling dijet invariant mass distribution, \mjj. 
New states decaying to two jets may introduce localized excesses in this distribution.
In QCD, due to $t$-channel poles in the cross-sections for the dominant scattering processes, most dijet production occurs at small angles $\theta^*$, defined as the polar 
angle in the dijet centre-of-mass frame.~\footnote{Since, experimentally, the two partons cannot be distinguished, $\theta^*$ is always taken between 0 and $\pi/2$ with respect to the beam.}
Many theories of BSM physics predict additional dijet production with a significant population of jets produced at large angles with respect to the beam; for reviews see Refs.~\cite{Harris:2011bh,BoelaertAkesson:2010}.
The search reported in this \publicationType exploits these generic features of BSM signals in an analysis of the \mjj and angular distributions.

\newcommand{\AtlasCoordFootnote}{
ATLAS uses a right-handed coordinate system with its origin at the nominal interaction point (IP) in the centre of the detector and the $z$-axis along the beam line.
The $x$-axis points from the IP to the centre of the LHC ring, and the $y$-axis points upwards.
Cylindrical coordinates $(r,\phi)$ are used in the transverse plane, $\phi$ being the azimuthal angle around the $z$-axis.
The pseudorapidity is defined in terms of the polar angle $\theta$ as $\eta = -\ln \tan(\theta/2)$.
It is equivalent to the rapidity for massless particles.

}

As is common, a rapidity $y = \ln{(E+p_z)/(E-p_z)}/2$ is defined for each of the outgoing partons, where $E$ is its energy and $p_z$ is the component of its momentum along the beam line.\footnote{\AtlasCoordFootnote\xspace}
Each incoming parton carries a fraction (Bjorken $x$) of the momentum of the proton. A momentum imbalance between the two partons boosts the centre-of-mass frame of the collision relative to the laboratory frame along the $z$ direction by $y_{\rm B} = \ln{(x_1/x_2)}/2 = (y_3+y_4)/2$, where $y_B$ is the rapidity of the boosted centre-of-mass frame, $x_1$ and $x_2$ are the fractions of the proton momentum carried by each parton and $y_3$ and $y_4$ are the rapidities of the outgoing partons in the detector frame.
Differences between two rapidities are invariant under such Lorentz boosts, hence the following function of the rapidity difference $\yStar = (y_3-y_4)/2$ between the two jets,

\begin{equation*}
  \label{eq:chidefinition}
  \chi = {\rm e}^{2|\yStar|} \sim \frac{1+\cos{\thetaStar}}{1-\cos{\thetaStar}} ,
\end{equation*}

\noindent is the same in the detector frame as in the partonic centre-of-mass frame.
In the centre-of-mass frame, the two partons have rapidity $\pm \yStar$.

The variable $\chi$ is constructed such that in the limit of massless parton scattering, and when only $t$-channel scattering contributes to the partonic cross-section, the angular distribution ${\rm d}N/{\rm d}\chi$ is approximately independent of $\chi$.
The measured shapes of the observed $dN/d\chi$ distributions differ from the parton-level distributions because the observed distributions convolve the parton-level distributions with non-uniform parton momentum distributions in $x_1$ and $x_2$. 
Restricting the range of two-parton invariant mass and placing an upper cut on $y_{\rm B}$ 
reduces these differences.

Prior searches of dijet distributions with lower-energy hadron collisions at the S$\mathrm{\bar{p}}$pS~\cite{Arnison:1983dk,Albajar:1988rs,Bagnaia1984283}, the Tevatron~\cite{CDF:2009DijetSearch,DZero:2009DijetAng}, and the LHC at $\sqrt{s}=7$--8~\TeV\xspace~\cite{ATLAS:2010bc,ATLAS:2010eza,CMS:2010DijetMass,CMS:2010centrality,CMS:2011DijetMassAngle,CMS:2011-1107-4771,Aad:2011aj,Aad:2011fq,ATLAS:2013eza,ATLAS:2015-1407-1376,CMS:2013-1302-4794,ATLAS:1504-00357} and recently at 13 TeV~\cite{Khachatryan:2015dcf}, did not find BSM phenomena.
This \publicationType presents an analysis of \integLumi of proton--proton collision LHC data at $\sqrt{s}=13$~\TeV\xspace recorded by the ATLAS detector, focusing on the distributions of \mjj and $\chi$ with methods based on those used by Refs.~\cite{ATLAS:2015-1407-1376,ATLAS:1504-00357}.

%% file: 02detector.tex
The ATLAS experiment~\cite{DetectorPaper} at the LHC is a multi-purpose particle detector with a forward-backward symmetric cylindrical geometry with layers of tracking, calorimeter, and muon detectors over nearly the entire solid angle around the $pp$ collision point.
The directions and energies of high-\pT{} hadronic jets are measured using silicon tracking detectors and straw tubes detecting transition radiation, 
finely segmented hadronic and electromagnetic calorimeters, and a muon spectrometer.
A steel\allowbreak/scintillator-tile calorimeter provides hadronic energy measurements for the pseudorapidity range $|\eta| <  1.7$.
A lead\allowbreak/liquid-argon (LAr) calorimeter provides electromagnetic (EM) energy measurements with higher granularity within the region $|\eta|< 3.2$.
The end-cap and forward regions are instrumented with LAr calorimeters for EM and hadronic energy measurements up to $|\eta| = 4.9$.
The first-level trigger is implemented in hardware and uses a subset of the detector information to reduce the accepted rate to \SI{100}{\kilo\hertz}.
This is followed by a software-based trigger that reduces the rate of events recorded to \SI{1}{\kilo\hertz}.

%% file: 03selection.tex
Collision events are recorded using a trigger requiring the presence of at least one jet reconstructed in the software-based trigger with a \pT of at least 360~\GeV.
Groups of contiguous calorimeter cells (topological clusters) are formed based 
on the significance of the energy deposit over calorimeter noise~\cite{ATLAS-LARG-PUB-2008-002}. 
Topological clusters are grouped into jets using the anti-$k_t$ algorithm~\cite{antikT,Cacciari:2006} with radius parameter $R=0.4$. 
Jet four-momenta are computed by summing over the topological clusters that constitute each jet, treating the energy of each cluster as a four-momentum with zero mass. 
The reconstruction efficiency for jets with \pT~above 20 GeV is 100\%. 
Jet calibrations derived from $\sqrt{s}=13~\TeV$~simulation, and collision data taken at $\sqrt{s}=8$~\TeV\ and $\sqrt{s}=13$~\TeV, are used to correct the jet energies and directions to those of the particles from the hard-scatter interaction.
This calibration procedure, described in Refs.~\cite{Aad:2015ina, ATLAS-CONF-2015-017,Alkire:2018215}, is improved by a data-derived correction to the relative calibration of jets in the central and the forward regions. The dijet mass resolution is 2.4\% and 2\%, for dijet masses of 2 and 5 TeV respectively. 
The jet energy scale uncertainty from 8~\TeV~data is complemented by systematic uncertainties covering the differences between 8~\TeV~and~13~\TeV~data. The total jet energy scale uncertainty is 1\% for central jets with \pT~of 500~\GeV, and 3\% for jets of~2~\TeV. 
Analysis of jet data at 13~\TeV~using the {\it in situ} techniques described in Ref.~\cite{ATLAS-CONF-2015-002} confirms the jet calibration and uncertainty estimates. Beyond the \pT\ range of the {\it in situ} techniques, for the quantities used to calibrate jets as well as other kinematic quantities, the data agree with simulation within quoted uncertainties.  

Events containing at least two jets are selected for offline analysis if the \pT of the leading and subleading jets is greater than 440~\GeV and 50~\GeV respectively.
This requirement ensures a trigger efficiency of at least 99.5\% for collisions with $|\yStar|<1.7$ and removes a negligible number of events from unbalanced
dijet events originating from additional interactions within the same bunch crossing or jet resolution tails.
Events are discarded from the search if any of the three leading jets with $\pT>$~50~\GeV~is compatible with non-collision background or calorimeter noise \cite{Makovec:2016323}.

%% file: 04simulation.tex
For this search, events from QCD processes are simulated with {\sc Pythia~8}~\cite{pythia8} using the A14~\cite{A14tune} set of tuned parameters for the underlying event and the leading-order NNPDF2.3~\cite{Ball:2012cx} parton distribution functions (PDFs).
The renormalization and factorization scales are set to the average \pT~of the two leading jets.
Detector effects are simulated using {\sc Geant4}~\cite{Agostinelli:2002hh} within the ATLAS software infrastructure~\cite{Aad:2010ah}.
The same software used to reconstruct data was also used to reconstruct simulated events.
The simulated events are used to predict the angular distribution from QCD processes and for qualitative comparisons to kinematic distributions in data. 

{\sc Pythia~8} calculations use matrix elements that are at leading order in the QCD coupling constant with simulation of higher-order contributions partially covered by the parton shower (PS) modelling. 
They also include modelling of hadronization effects.
The distributions of events predicted by {\sc Pythia~8} are reweighted to the next-to-leading-order (NLO) predictions of {\sc NLOJET}++~\cite{NLOJETPP1, NLOJETPP2, NLOJETPP3} using mass- and $\chi$-dependent correction factors defined as in Ref.~\cite{ATLAS:1504-00357}.
The correction factors modify the shape of the angular distributions at the level of 15\% at low values of $\chi$ and high values of \mjj. The correction is 5\% or less at the highest values of $\chi$. 
The {\sc Pythia~8} predictions also omit electroweak effects. 
These are included as additional mass- and $\chi$-dependent correction factors~\cite{Sudakov} 
that are unity at low \mjj and differ from unity by up to 3\% in the $\mjj > 3.4$~\TeV~region.

BSM signal samples of excited quarks~\cite{Baur:1987ga,Baur:1989kv}, new heavy vector bosons~\cite{Altarelli:1989ff,LHCDMF:2015,Chala:2015ama}, quantum black holes~\cite{Gingrich:2009hj,Calmet:2008dg,Dai:2007ki} and contact interactions~\cite{CI1, CI2, CI3} are simulated and reconstructed using the same procedure as for QCD processes. The models and the parameters chosen for generation are described in Section~\ref{sec:bsm}.

%% file: 05resonance.tex
\begin{figure}[h!]
  \centering
  \includegraphics[width=1.00\columnwidth]{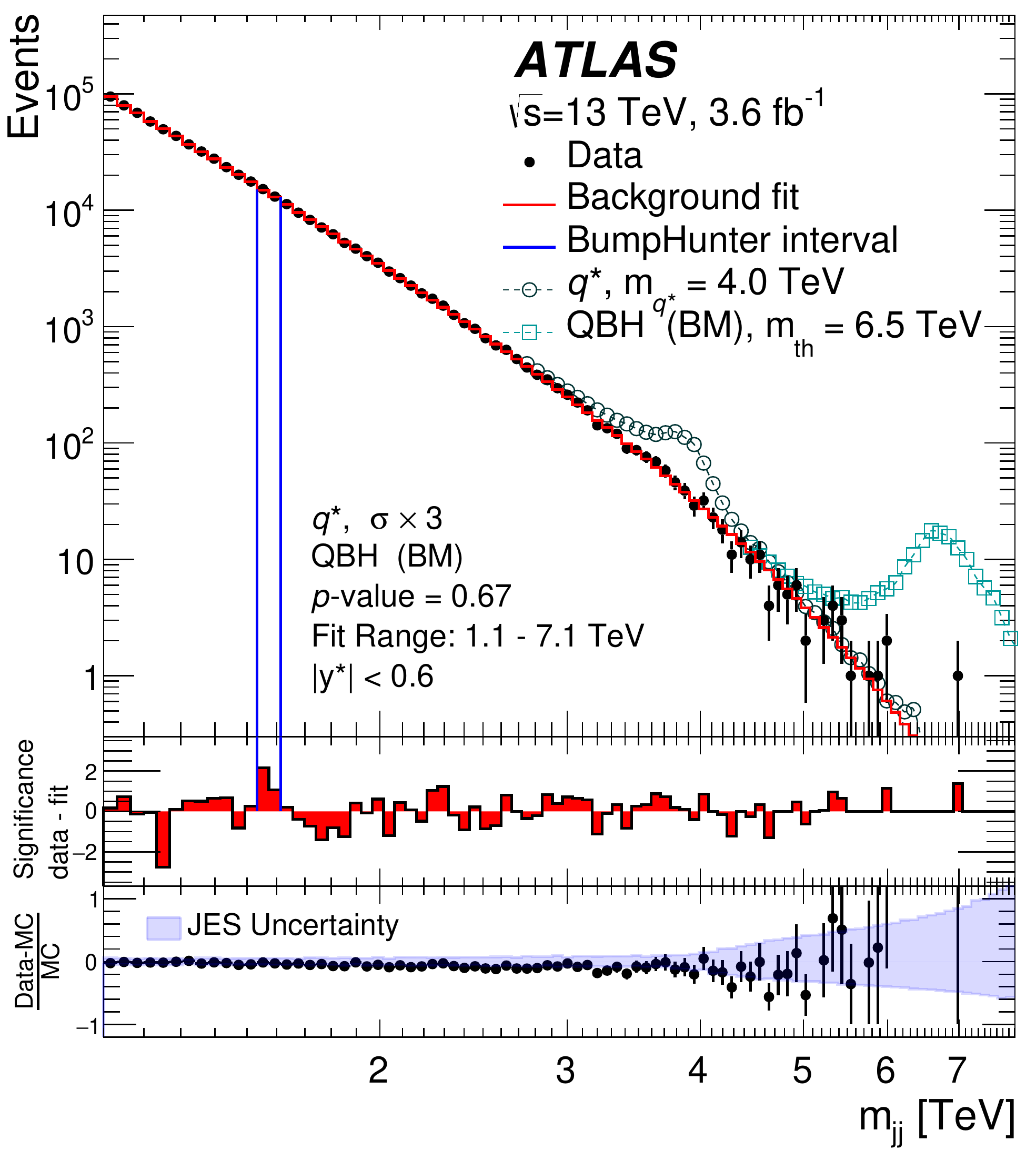}
  \caption{The reconstructed dijet mass distribution (filled points) for events with $|y^*|~<~0.6$ and
    $\pT>440$~$(50)$~$\GeV\xspace$ for the leading~(subleading) jets.
    The solid line depicts the fit to Eq.~(\ref{eq:backgroundfit}), as discussed in the text.
    Predictions for an excited quark and a quantum black hole signal predicted by the {\sc BlackMax} generator (QBH BM) are shown above the fit, normalized to the predicted cross-section. 
    The vertical lines indicate the most discrepant interval identified by the {\sc BumpHunter} algorithm, for which the $p$-value is stated in the figure. 
    The middle panel shows the bin-by-bin significances of the data--fit differences, considering only statistical uncertainties.
    The lower panel shows the relative differences between the data and the prediction of {\sc Pythia~8} simulation of QCD processes, corrected for NLO and electroweak effects, and is shown purely for comparison.
    The shaded band denotes the experimental uncertainty in the jet energy scale calibration.
    }
  \label{fig:one}
\end{figure}

The \mjj distribution of events with $|\yStar| < 0.6$ ($\chi<3.3$) is analysed for evidence of contributions from resonant BSM phenomena.
The requirement on $|\yStar|$ reduces the background from QCD processes.
To avoid kinematic bias from the \yStar\ and \pt\ selections described above, the analysis is confined to $\mjj> 1.1$~\TeV\xspace. 

Fig.~\ref{fig:one} shows the observed \mjj distribution for the resonance selection, overlaid with examples of the signals described in Section~\ref{sec:bsm}. 
The bin widths are chosen to approximate the \mjj resolution as derived from the simulation of QCD processes, and therefore widen as the mass increases.
The largest value of \mjj measured is \HighestDijetMassResonance.

To estimate the SM background, the ansatz,
\begin{equation}
  \label{eq:backgroundfit}
   f(z)=p_{1}(1-z)^{p_{2}}z^{p_{3}}, 
\end{equation}
where $z \equiv \mjj/\sqrt{s}$, is fit to the $m_{jj}$ distribution in Fig.~\ref{fig:one} to obtain the parameters $p_{i}$. The fit range is \resonanceFitRange.
CDF, CMS, and ATLAS dijet searches such as those described in Refs.~\cite{CDF:2009DijetSearch,ATLAS:2010bc,CMS:2011-1107-4771,Aad:2011aj,ATLAS:2015-1407-1376} have found that expressions similar to Eq.~(\ref{eq:backgroundfit}) describe dijet mass distributions observed at lower collision energies.
The ansatz also describes leading-order and next-to-leading order simulations of QCD dijet production at $\sqrt{s}=13$~\TeV.
A log-likelihood-ratio statistic employing Wilks’s theorem~\cite{wilks1938} was used to determine if the background estimation 
would be significantly improved by an additional degree of freedom.
With the current dataset, Eq.~(\ref{eq:backgroundfit}) was found to be sufficient.

Fig.~\ref{fig:one} also shows the result of the fit. 
The fit describes the observed data with \resonanceBackgroundFitQualityStatement. 
The middle panel of the figure shows the significances of bin-by-bin differences between the data and the fit. 
These Gaussian significances are calculated from the Poisson probability, considering only statistical uncertainties. 
The lower panel compares the data to the prediction of {\sc Pythia~8} simulation of QCD processes, corrected for NLO and electroweak effects. Even though it is not used in the analysis of the \mjj distribution, the simulation is shown to be in good agreement with the data.

The uncertainty in values of the parameters in Eq.~(\ref{eq:backgroundfit}) is evaluated by fitting them to pseudo-data drawn via Poisson fluctuations around the fitted background model.
The uncertainty in the prediction in each \mjj bin is taken to be the root mean square of the function value for all pseudo-experiments in that bin.
To estimate an uncertainty due to the choice of the background parameterization, a parameterization with one additional degree of freedom, $z^{p_4 \log{z}}$, is compared to the nominal ansatz, and the difference is taken as an uncertainty.
The prediction of the \mjj\xspace distribution does not involve simulated collisions and thus 
is not affected by theoretical or experimental uncertainties.

The statistical significance of any localized excess in the  \mjj\xspace distribution is quantified using the {\sc BumpHunter} algorithm~\cite{Aaltonen:2008vt, Choudalakis:2011bh}.
The algorithm compares the binned \mjj distribution of the data to the fitted background estimate, considering contiguous mass intervals in all possible locations, from a width of two bins to a width of half of the distribution.
For each interval in the scan, it computes the significance of any excess found.
The algorithm identifies the interval \bumpHunterSignalInterval, indicated by the two vertical lines in Fig.~\ref{fig:one}, as the most discrepant interval.
The statistical significance of this outcome is evaluated using the ensemble of possible outcomes across all intervals scanned, by applying the algorithm to many pseudo-data samples drawn randomly from the background fit. 
Without including systematic uncertainties, the probability that fluctuations of the background model would produce an excess at least as significant as the one observed in the data, anywhere in the distribution, is \bumpHunterPValue.
Thus, there is no evidence of a localized contribution to the mass distribution from BSM phenomena.

%% file: 06angular.tex
\begin{figure}[h!]
  \centering 
  \includegraphics[width=1.00\columnwidth]{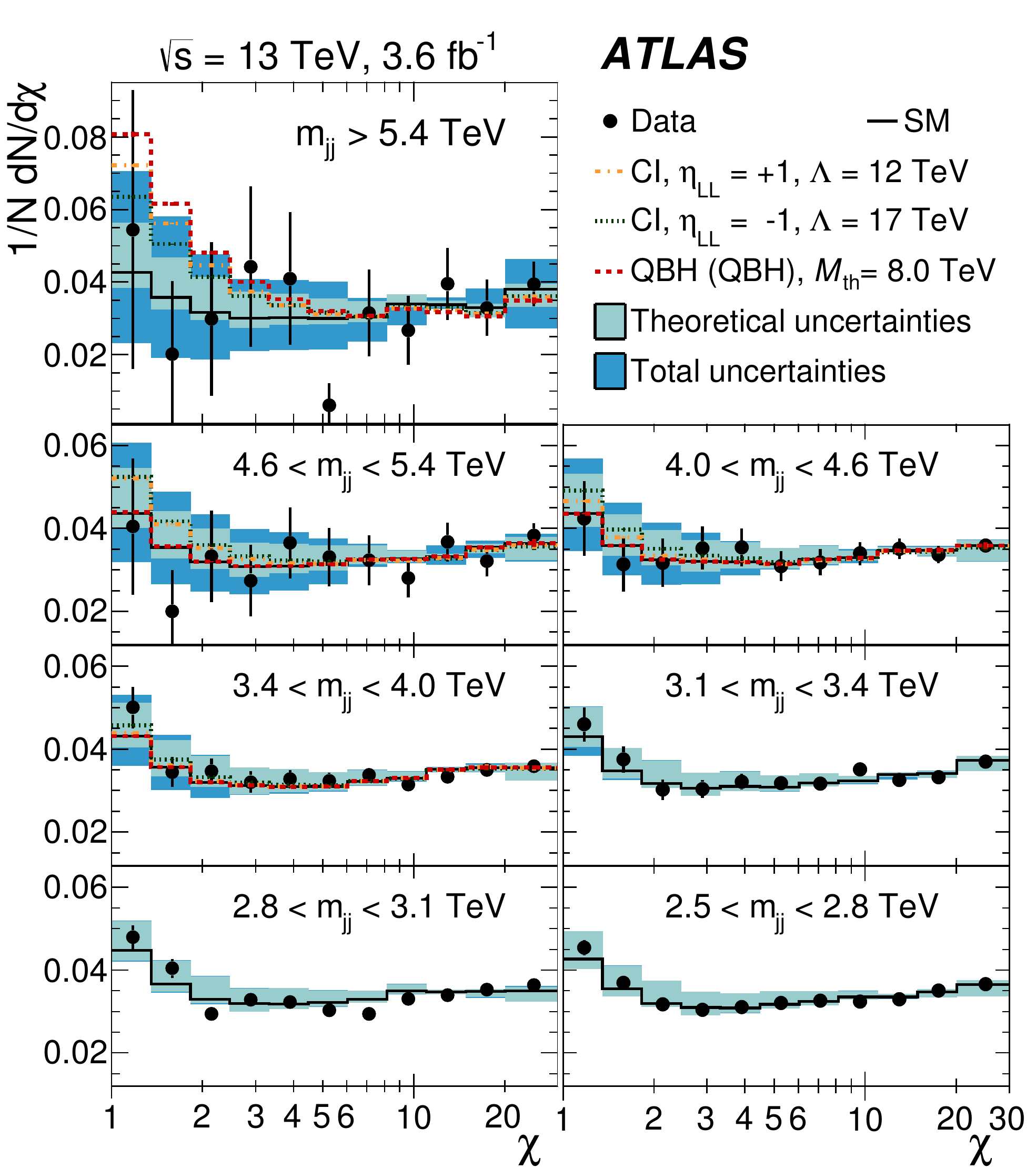}
  \caption{Reconstructed distributions of the dijet angular variable $\chi$ in
    different regions of the dijet invariant mass \mjj for events with
    $|\yStar|<1.7$, $|y_{\rm B}| < 1.1$ and $\pT>440$~$(50)$~$\GeV\xspace$
    for the leading~(subleading) jets. Shown are the data (points),
    corrected NLO predictions (solid lines), and examples of
    the contact interaction (CI) and quantum black hole (QBH) signals discussed in the text.
    The theoretical uncertainties and the
    total theoretical and experimental uncertainties in the predictions are displayed as
    shaded bands around the SM prediction.}
  \label{fig:two}
\end{figure}

The ${\rm d}N/{\rm d}\chi$ (angular) distributions of events with $|\yStar| < 1.7$ (i.e. $\chi < 30.0$) and $|y_{\rm B}| < 1.1$ are also analysed for contributions from BSM signals.
Fig.~\ref{fig:two} shows the angular distributions of the data in different \mjj ranges, the SM prediction for the shape of the angular distributions, and examples of the signals described in Section~\ref{sec:bsm}. 
The data with $\mjj< 2.5$~\TeV\xspace are discarded to remove bias from the kinematic selections described earlier.
The highest \mjj measured is \HighestDijetMassAngular.
The SM prediction is obtained from simulation, as described in Section~\ref{sec:simulation}.
In the analysis, the prediction in each \mjj range is normalized to match the integral of the data in that range.

Theoretical uncertainties in simulations of the angular distributions from QCD processes are estimated as described in Ref.~\cite{ATLAS:1504-00357}.
The effect on the QCD prediction of varying the PDFs is estimated using {\sc NLOJET}++ with three different PDF sets: CT10~\cite{Lai:2010vv}, MSTW2008~\cite{Martin:2009iq} and NNPDF23~\cite{Ball:2012cx}.
As the choice of PDF largely affects the total cross-section rather than the shape of the $\chi$ distributions, these uncertainties are negligible ($<1\%$).
The uncertainty due to the choice of renormalization and factorization scales was estimated using {\sc NLOJET}++ by varying each independently up and down by a factor two, excluding opposite variations. 
The resulting uncertainty, taken as the envelope of the variations in the normalized $\chi$ distributions, depends on both \mjj and $\chi$, rising to 20\% at the smallest $\chi$ values at high \mjj values.
The statistical uncertainty of the simulated NLO corrections is less than 1\%.
The dominant experimental uncertainty in the predictions of the $\chi$ distributions is the jet energy scale uncertainty, with an impact of at most 25\% at high \mjj values. 
The uncertainty in the jet energy resolution has negligible impact.
The theoretical uncertainties and the total uncertainties are displayed as shaded bands around the prediction.

The $CL_{s}$ technique~\cite{Read:2002hq,Junk:1999kv} is used to test the compatibility of the $\chi$ distribution 
with the SM prediction and with the BSM signals discussed in Section~\ref{sec:bsm}, 
using a combined fit in four coarse $\mjj$ bins covering $\mjj > 3.4$~\TeV. 
No significant deviation of the data from the background-only hypothesis is observed, with a~$CL_{b}$~of 0.35.

%% file: 07signals.tex
The data are used to constrain several of the many BSM models that predict dijet excesses. Quantum black holes, excited quarks, and $W^\prime$ and $Z^\prime$ bosons would produce peaks in the \mjj\xspace distribution. Contact interactions would introduce smooth changes in the high-mass tail of the $\mjj$ distribution that could be detected in the analysis of the $\chi$ distributions. 
The signal models are simulated using the parton-level generators indicated below, in an identical manner to QCD processes, using the same PDFs and parameters for non-perturbative effects, except where noted otherwise.

The LHC could produce black holes with masses at or above the fundamental scale of gravity, $M_{\rm D}$, if that scale is lowered to a few \TeV\xspace by the existence of extra spatial dimensions~\cite{Giddings:2001bu,Dimopoulos:2001hw,RandallMeade,Feng:2004,Calmet:2008dg,Gingrich:2009hj,BoelaertAkesson:2010}. 
High-multiplicity final states from thermalizing black holes are explored at $\sqrt{s} =13$~TeV by ATLAS in Ref.~\cite{ATLAS:1512-futureMultijet}.
This analysis explores quantum black holes (QBHs), which would be produced near $M_{\rm D}$ and decay into a few particles rather than high-multiplicity final states \cite{Gingrich:2009hj,Calmet:2008dg,Dai:2007ki,Frost:2009cf}, appearing in the \mjj distribution as an excess localized at the 
threshold mass for the quantum black hole production, \mthreshold. 
Here, production and decay to two jets is simulated using the {\sc QBH} generator~\cite{Gingrich:2009da} or the {\sc BlackMax} generator~\cite{Dai:2007ki},\footnote{Black holes decay thermally to non-rotating QBH in {\sc BlackMax}, while the decay products of the {\sc QBH} generator are dictated by local gauge symmetries of the SM.} assuming an Arkani-Hamed\discretionary{--}{}{--}Dimopoulous\discretionary{--}{}{--}Dvali (ADD) scenario~\cite{ArkaniHamed:1998rs,Antoniadis:1998ig} with $M_{\rm D}=\mthreshold$ and a number of extra dimensions $n = 6$, as in Ref.~\cite{ATLAS:2015-1407-1376}, and a Randall--Sundrum scenario (RS1)~\cite{Randall:1999ee} with $n = 1$ using the {\sc QBH} generator. 
In these models, the branching ratio to dijets is greater than 96\%.
The acceptance times efficiency of the resonance (angular) selection for a quantum black hole with a threshold mass of 6.5~\TeV\xspace is 53\% (92\%) for both generators. 
The PDFs used are CTEQ6L1~\cite{Pumplin:2002vw}.

Excited quarks ($q^*$)~\cite{Baur:1987ga,Baur:1989kv} are predicted in models of compositeness and are a benchmark for quark--gluon resonances~\cite{ATLAS:2010bc,Aad:2011fq,Aad:2011aj,ATLAS:2010eza}.
The $q^*$ model is simulated with {\sc Pythia~8}, assuming spin-1/2 excited quarks with coupling constants the same as for SM quarks.
As in Ref.~\cite{Baur:1989kv}, the compositeness scale is set equal to the excited quark mass, $m_{q^*}$, and the SU(3), SU(2), and U(1) coupling multipliers $f_{\rm s} = f = f^{\prime} = 1$. 
The renormalization and factorization scales are set to the average \pt\ of the two leading jets. 
In the simulation, only the decay of the excited quark to a gluon and an up- or down-type quark is modelled; this corresponds to a branching ratio of 85\%. 
Before parton shower effects are taken into account,
the intrinsic width of the $q^*$ signals is comparable to the detector resolution. 
The resonance selection acceptance times efficiency for a $q^*$ with a mass of 4~\TeV\xspace is 58\%.

Additional spin-1 $W^\prime$ and $Z^\prime$ bosons often arise in the symmetry breaking of extended gauge theories.
A $W^\prime$ model~\cite{Altarelli:1989ff} with~$V$--$A$~SM couplings 
and a corresponding branching ratio to dijets of 75\% is considered.
In this analysis, events are simulated in {\sc Pythia~8} and decays 
are restricted to quark--antiquark pairs with all six quark flavours included.
Events including top decays were not removed from the analysis, resulting
in conservative limits.
A leptophobic $Z^\prime$ model \cite{LHCDMF:2015} is also simulated, with matrix elements calculated in {\sc MadGraph 5}~\cite{Alwall:2014hca} and parton showering performed in {\sc Pythia~8}.
The $Z^\prime$ model assumes axial-vector couplings to all SM quarks and to a Dirac fermion dark matter candidate.
No interference with the SM is simulated for either the $W^\prime$ or the $Z^\prime$ model and decays involving top quarks are included. 
The $Z^\prime$ model considered follows a scenario~\cite{Chala:2015ama} where its decays to dark matter are negligible, 
hence the dijet production rate and resonance width depend only on the coupling to quarks, $g_q$, and the mass of the resonance $m_{Z^\prime}$.
Before parton shower effects are considered,
the intrinsic width of the  $W^\prime$ and $Z^\prime$ signals range from 0.05\% for a $Z^\prime$ with a mass of 1.5~\TeV\xspace and $g_q=0.1$ to
10\% for a $Z^\prime$ with a mass of 3.5~\TeV\xspace and $g_q=0.5$. 
The resonance selection acceptance times efficiency for a mass of 3~\TeV\xspace is 40\% for the $W^\prime$ model and 47\% for the $Z^\prime$ model with $g_q=0.2$.

Results are also provided as limits on the cross-section times acceptance times branching ratio to two jets,
$\sigma\times A\times\text{BR}$, of a hypothetical signal that produces a Gaussian contribution to the observed \mjj distribution. 
For sufficiently narrow resonances, these results may be used to set limits in BSM models beyond those considered explicitly in this \publicationType. 
These limits should be used when PDF and non-perturbative effects can be safely truncated or neglected and, after applying the resonance selection, the reconstructed \mjj distribution predicted by the model approaches a Gaussian distribution.
Predicted BSM signals with an intrinsic width much smaller than 5\% should be compared to the limit curve for width equal to the experimental resolution.
Predicted signals with larger widths should be compared with the limit that corresponds most closely to the width of the Gaussian contribution predicted by the model.
More instructions can be found in Appendix A of Ref.~\cite{ATLAS:2015-1407-1376}.

For all signals described above, the following systematic uncertainties are included in the limit setting: jet energy scale, PDF and uncertainties due to higher-order corrections, luminosity, and statistical uncertainties of the simulated events. The jet energy uncertainty is up to 10\%. 
On average, the PDF uncertainty affects the angular distributions by 1\%.
The uncertainty in the integrated luminosity is $\pm 9$\%. It is derived, following a method similar to that detailed in Ref.~\cite{Aad:2013ucp}, from a preliminary calibration of the luminosity scale using a pair of $x$--$y$ beam-separation scans performed in June 2015.

The dijet distributions can also be modified by new mediating particles with a mass much higher than can be probed directly. A four-fermion effective field theory (contact interaction)~\cite{CI1, CI2, CI3} characterized by a single energy scale $\Lambda$ can then be used to describe these effects:
\begin{eqnarray} 
L_{qq} =  \frac{2\pi}{\Lambda^2} [&\eta_{\mathrm{LL}}(\bar{q}_{\rm L}\gamma^\mu q_{\rm L})(\bar{q}_{\rm L}\gamma_\mu q_{\rm L}) \nonumber \\ 
&+ \eta_{\mathrm{RR}}(\bar{q}_{\rm R}\gamma^\mu q_{\rm R})(\bar{q}_{\rm R}\gamma_\mu q_{\rm R}) \nonumber \\
&+ 2\eta_{\mathrm{RL}}(\bar{q}_{\rm R}\gamma^\mu  q_{\rm R})(\bar{q}_{\rm L}\gamma_\mu q_{\rm L}) ],
\end{eqnarray}
\noindent
where the quark fields have L and R chiral projections and the coefficients $\eta_{\mathrm{LL}}$, $\eta_{\mathrm{RR}}$, and $\eta_{\mathrm{RL}}$ turn on and off various interactions. 
Contact interactions with a non-zero left-chiral colour-singlet coupling ($\eta_{\mathrm{LL}}= \pm 1$, $\eta_{\mathrm{RL}}=\eta_{\mathrm{RR}}= 0$) are simulated using {\sc Pythia~8}. 
This type of coupling is chosen because its angular distributions are representative of 
those of other BSM models.
Interference of the signal model with the SM process $q\bar{q}~\rightarrow~q\bar{q}$ is included.
Events are simulated for both constructive and destructive interference with $\Lambda = 7$~\TeV\xspace. 
From this sample, the angular distributions for other values of $\Lambda$ are obtained using the fact that the interference term is proportional to $1/\Lambda^2$ and the pure contact-interaction cross-section is proportional to $1/\Lambda^4$.
The {\sc Pythia~8} signal prediction is reweighted to the NLO cross-sections provided by CIJET~\cite{CIJET}.
Uncertainties in the prediction of the angular distributions for contact-interaction signals are obtained in the same manner as for QCD processes.

%% file: 09limits.tex
\begin{figure}[h!]
  \centering
  \includegraphics[width=1.00\columnwidth]{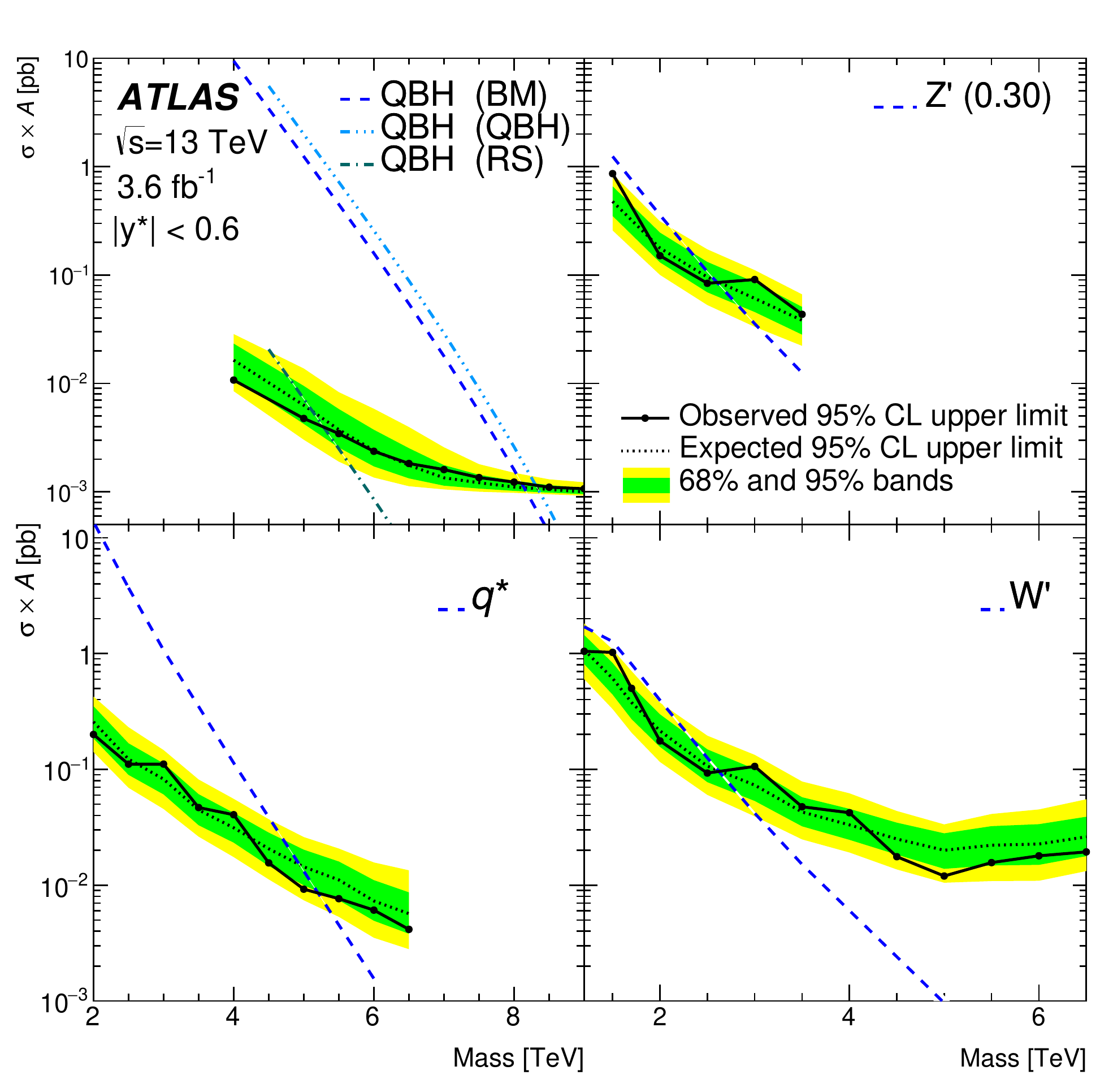}
  \caption{The 95\% credibility-level upper limits obtained from the \mjj distribution on cross-section, $\sigma$, times acceptance, $A$, for the models described in the text. Clockwise from top left:
Quantum black holes with $n=6$ generated with {\sc BlackMax} (QBH~(BM)), and with $n=6$ and $n=1$ with {\sc QBH} (denoted by QBH~(QBH) and QBH~(RS), respectively), 
$Z^{\prime}$ with $g_q=0.3$, $W^{\prime}$, and $q^\star$.  }
  \label{fig:limitsModels}
\end{figure}
\begin{figure}[h!]
	\centering
	\includegraphics[width=1.00\columnwidth]{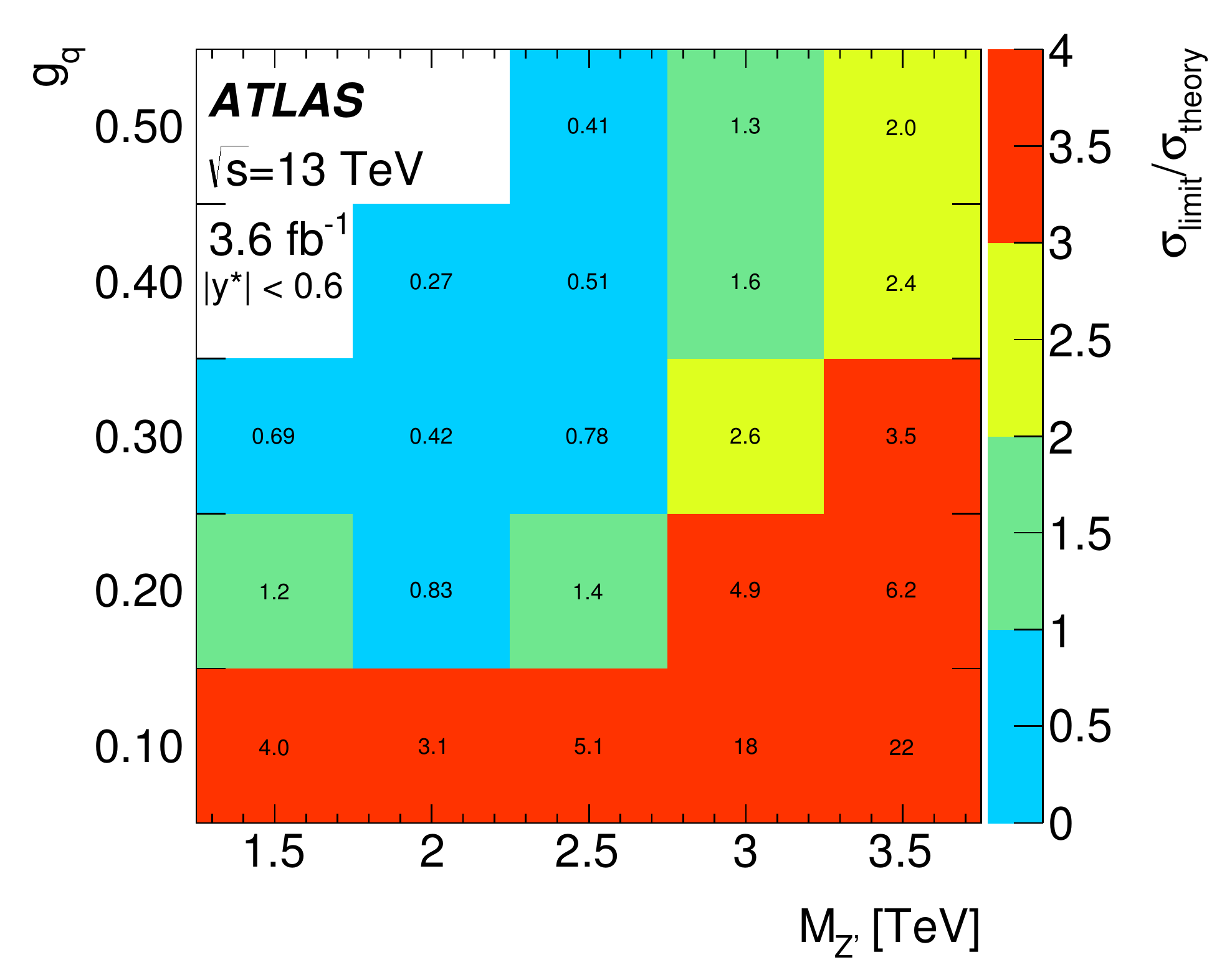}
	\caption{The ratio of 95\% credibility-level upper limits to predicted cross-sections with respect to the $Z^{\prime}$ model predictions described in the text, as a function of the coupling to quarks, $g_q$, and the mass, $M_{Z^\prime}$, obtained from the \mjj distribution. Since for a given mass higher couplings have higher cross sections and would therefore be excluded if lower couplings are excluded, the limits are not calculated in the white area. }
	\label{fig:zprimeTwoD}
\end{figure}

Starting from the \mjj distribution obtained with the resonance selection, a Bayesian method~\cite{Aad:2011aj} is applied to the data and simulation of signals at a series of discrete masses to set 95\% credibility-level upper limits on the cross-section times acceptance for the signals described above.
The method uses a constant prior for signal cross-section and Gaussian priors for nuisance parameters corresponding to systematic uncertainties.
The expected limits are calculated using pseudo-experiments generated from the maximum-likelihood values for parameters of the background-only model in Eq.~(\ref{eq:backgroundfit}) using the full systematic uncertainties in both the signal and background models.
The limit is interpolated logarithmically between the discrete masses to create curves continuous in signal mass.
The mass limits for each of those models are shown in Figs.~\ref{fig:limitsModels} and \ref{fig:zprimeTwoD} and Table~\ref{tab:Limits}.
No uncertainty is included for the cross-section of the signals considered. 

\begin{table}[!h]
	\centering
	\resizebox{\columnwidth}{!}{
		\begin{tabular}{| l |c c c|}
			\hline
			\multicolumn{1}{|c|}{Model} & \multicolumn{3}{c|}{95\% CL Exclusion limit} \\
      & Run 1 Observed  &  Observed 13 TeV & Expected 13 TeV \\
			\hline
      Quantum black holes, ADD & \multirow{2}{*}{\blackMaxLimitMjjRunI } &	\multirow{2}{*}{\blackMaxLimitMjj } &\multirow{2}{*}{\blackMaxLimitMjjExpected } \\
			({\sc BlackMax} generator)     & &  &       \\              

      Quantum black holes, ADD & 	\multirow{2}{*}{\qbhLimitMjjRunI } & \multirow{2}{*}{\qbhLimitMjj } &\multirow{2}{*}{\qbhLimitMjjExpected } \\
			({\sc QBH} generator)     & & &        \\              

      Quantum black holes, RS & 	\multirow{2}{*}{--} & \multirow{2}{*}{\RSLimitMjj } &\multirow{2}{*}{\RSLimitMjjExpected } \\
			({\sc QBH} generator)     & &  &       \\              
			\hline
			
      Excited quark & \qstarLimitMjjRunI & \qstarLimitMjj  &   \qstarLimitMjjExpected  \\
			\hline

      $W'$  & \wprimeLimitMjjRunI & \wprimeLimitMjjLow  &   \wprimeLimitMjjExpected  \\
			\hline
                                               
      Contact interactions ($\eta_{LL}=$+1) & \LambdaLimitPlusChiRunI & \LambdaLimitPlusChi & \LambdaLimitPlusChiExpected\\
      Contact interactions ($\eta_{LL}=$$-1$) & \LambdaLimitMinusChiRunI & \LambdaLimitMinusChi & \LambdaLimitMinusChiExpected \\
			\hline

			\hline
		\end{tabular}
	}
	\caption{The 95\% credibility-level lower limit on the mass of quantum black holes, $W'$ models and excited quarks from the resonance selection, and the 95\% confidence-level lower limit on the scale of contact interactions for constructive ($\eta_{LL}=-1$) and destructive ($\eta_{LL}=+1$) from the angular selection. Limits on the $Z^\prime$ model are provided in Fig.~\ref{fig:zprimeTwoD}. For comparison between the results from the two selections, the corresponding limit on quantum black holes for the angular selection is \qbhLimitChi for the {\sc QBH} $n=6$ model. The Run 1 limits shown above were obtained in Refs.~\cite{ATLAS:2015-1407-1376,ATLAS:1504-00357}.}
	\label{tab:Limits}
\end{table}

\begin{figure}[h!]
  \centering
  \includegraphics[width=1.00\columnwidth]{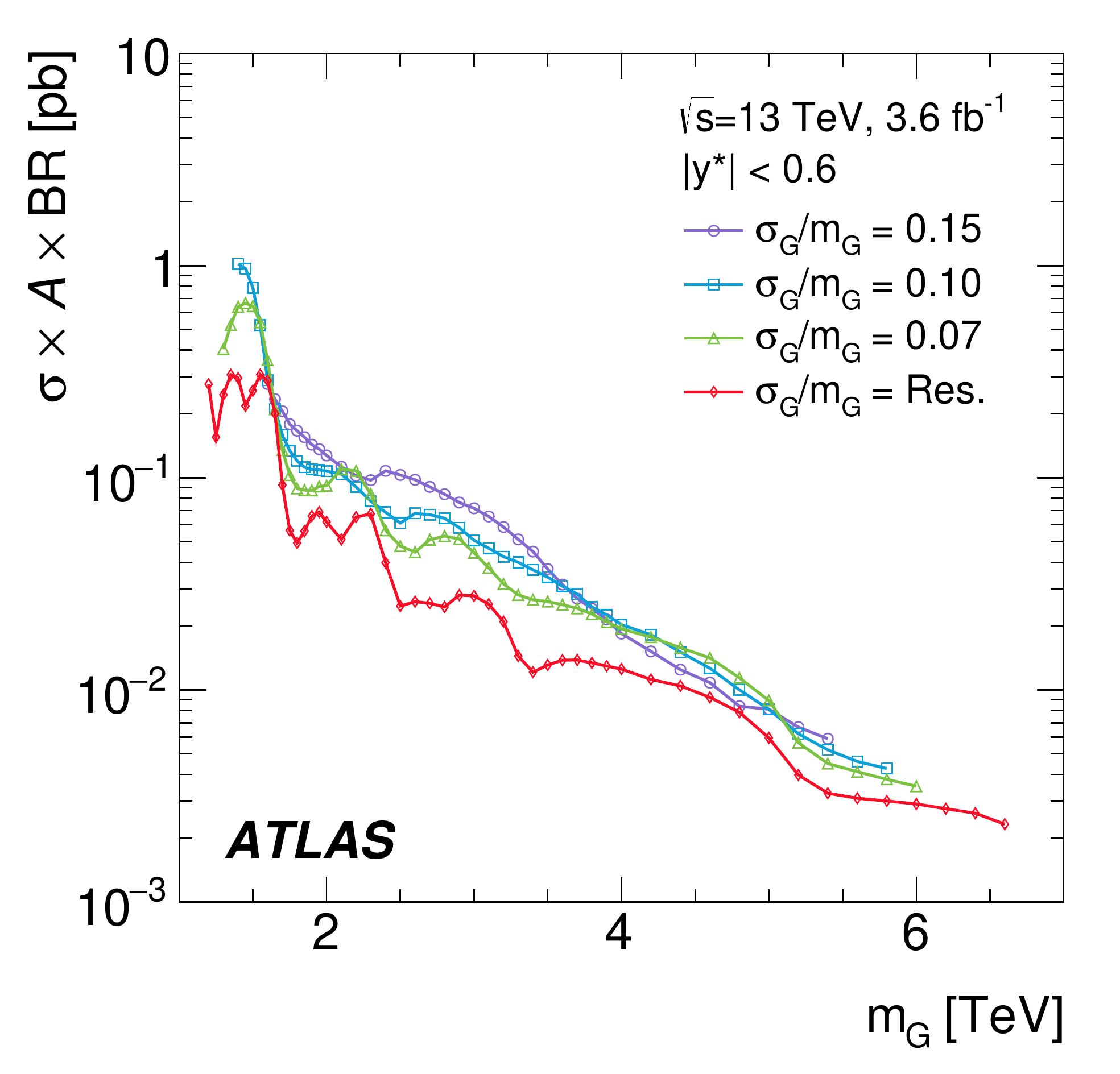}
  \caption{The 95\% credibility-level upper limits obtained from the \mjj distribution on cross-section times acceptance times branching ratio to two jets, $\sigma\times A\times\text{BR}$, for a hypothetical signal with a cross-section $\sigma_G$ that produces a Gaussian contribution to the observed \mjj distribution, as a function of the mean mass of the Gaussian distribution, $m_G$. Limits are obtained for four different widths, from a width equal to the detector mass resolution (``Res.''), 3\%--2\% depending on \mjj probed, to 15\% of the mean of the Gaussian mass distribution.}
  \label{fig:four}
\end{figure}

Fig.~\ref{fig:four} shows limits on the Gaussian contributions to the observed \mjj distribution obtained for a mean mass $m_G$ and four different widths, from a width equal to the detector mass resolution to a width of 15\% of the mean of the Gaussian mass distribution. 
Limits are set only when $m_G$ is within 1.1~\TeV--\HighestDijetMassResonance\xspace~and separated by at least twice the width of the Gaussian from the endpoints of this range.
Intrinsically narrow resonances with effective cross-sections exceeding values ranging \gaussianRangeStatement are excluded.
As the width increases, the expected signal contribution is distributed across more bins.
Therefore wider signals are affected less than narrower signals by statistical fluctuations of the data in a single bin.

\begin{figure}[!htb]
  \centering
  \includegraphics[width=1.00\columnwidth]{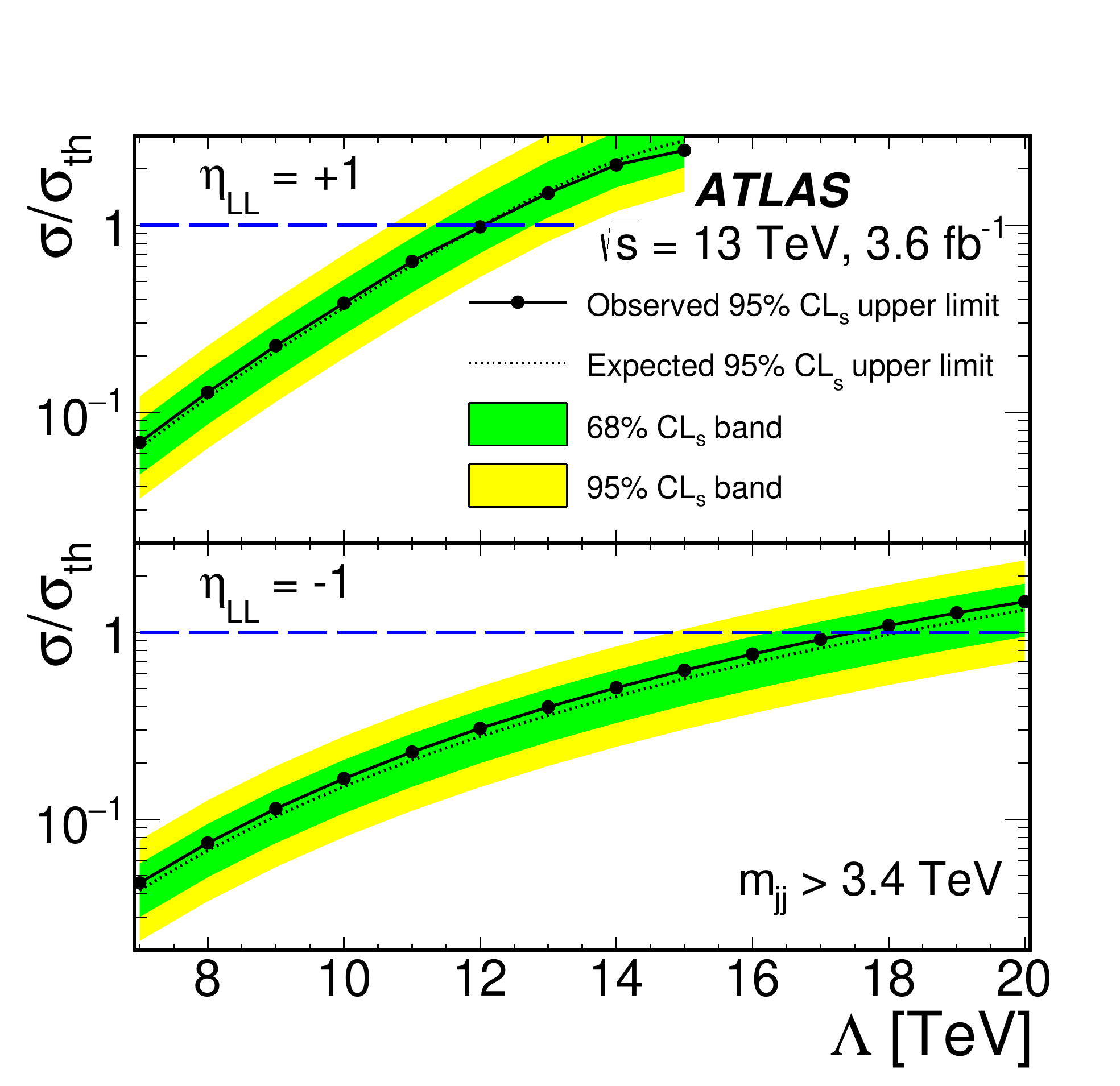}
  \caption{Ratio of the observed and expected 95\% confidence-level upper limits on the cross-section in the contact interaction model to the predicted cross-section $\sigma/\sigma_{\text{th}}$ as a function of compositeness scale $\Lambda$, for (top) destructive and (bottom) constructive interference with QCD processes. The crossing of the observed and expected 95\% confidence-level lines with the line at signal strength of one indicates observed and expected lower limits on $\Lambda$, respectively.}
  \label{fig:six}
\end{figure}

Starting from the $\chi$ distribution obtained with the angular selection, the $CL_{s}$ is calculated for signal contributions from contact interactions and quantum black holes, using the background predicted by the SM simulations as the null hypothesis.
The asymptotic approximation ~\cite{Cowan:2010js} of a profile likelihood ratio is used to set 95\% confidence-level limits in the contact interaction and quantum black hole models.
A combined fit is performed on the four highest-$m_{jj}$ regions of Fig.~\ref{fig:two}.
The correlation of the systematic uncertainties between the regions is taken into account
and the maximum likelihood values of the nuisance parameters do not differ significantly from the expectation.
The validity of the asymptotic approximation was confirmed using toy simulations.
The bounds on contact interactions are shown in Fig.~\ref{fig:six} and in Table~\ref{tab:Limits}.
Limits obtained from the angular distributions on quantum black hole signals are similar to the limits obtained from the $\mjj$\xspace distribution.

%% file: 10conclusions.tex
No evidence of phenomena beyond the Standard Model was uncovered in this search using dijet events in \integLumi of proton--proton collisions with a centre-of-mass energy of $\sqrt{s}=13$~TeV recorded by the ATLAS detector at the Large Hadron Collider.
The dijet invariant mass distribution exhibits no significant local excesses above a data-derived estimate of the smoothly falling distribution predicted by the Standard Model.
The dijet angular distributions also agree with a Monte Carlo simulation of the SM.
With the resonance selection, the analysis excludes at 95\% credibility level several types of signals, as predicted by models of quantum black holes, excited quarks, $W^\prime$ and $Z^\prime$ bosons. 
It also sets 95\% credibility-level upper limits on the cross-section for new processes that would produce a Gaussian contribution to the dijet mass distribution. 
It excludes Gaussian contributions if the effective cross-section exceeds values ranging \gaussianRangeStatement. 
With the angular selection, 95\% confidence-level lower limits are set on the compositeness scale of contact interactions at \LambdaLimitPlusChi~(\LambdaLimitMinusChi) for destructive (constructive) interference between the new interaction and QCD processes. These results significantly extend the ATLAS limits obtained from 8 TeV data.

%% file: acknowledgements/Acknowledgements.tex

We thank CERN for the very successful operation of the LHC, as well as the
support staff from our institutions without whom ATLAS could not be
operated efficiently.

We acknowledge the support of ANPCyT, Argentina; YerPhI, Armenia; ARC, Australia; BMWFW and FWF, Austria; ANAS, Azerbaijan; SSTC, Belarus; CNPq and FAPESP, Brazil; NSERC, NRC and CFI, Canada; CERN; CONICYT, Chile; CAS, MOST and NSFC, China; COLCIENCIAS, Colombia; MSMT CR, MPO CR and VSC CR, Czech Republic; DNRF, DNSRC and Lundbeck Foundation, Denmark; IN2P3-CNRS, CEA-DSM/IRFU, France; GNSF, Georgia; BMBF, HGF, and MPG, Germany; GSRT, Greece; RGC, Hong Kong SAR, China; ISF, I-CORE and Benoziyo Center, Israel; INFN, Italy; MEXT and JSPS, Japan; CNRST, Morocco; FOM and NWO, Netherlands; RCN, Norway; MNiSW and NCN, Poland; FCT, Portugal; MNE/IFA, Romania; MES of Russia and NRC KI, Russian Federation; JINR; MESTD, Serbia; MSSR, Slovakia; ARRS and MIZ\v{S}, Slovenia; DST/NRF, South Africa; MINECO, Spain; SRC and Wallenberg Foundation, Sweden; SERI, SNSF and Cantons of Bern and Geneva, Switzerland; MOST, Taiwan; TAEK, Turkey; STFC, United Kingdom; DOE and NSF, United States of America. In addition, individual groups and members have received support from BCKDF, the Canada Council, CANARIE, CRC, Compute Canada, FQRNT, and the Ontario Innovation Trust, Canada; EPLANET, ERC, FP7, Horizon 2020 and Marie Skłodowska-Curie Actions, European Union; Investissements d'Avenir Labex and Idex, ANR, Region Auvergne and Fondation Partager le Savoir, France; DFG and AvH Foundation, Germany; Herakleitos, Thales and Aristeia programmes co-financed by EU-ESF and the Greek NSRF; BSF, GIF and Minerva, Israel; BRF, Norway; the Royal Society and Leverhulme Trust, United Kingdom.

The crucial computing support from all WLCG partners is acknowledged
gratefully, in particular from CERN and the ATLAS Tier-1 facilities at
TRIUMF (Canada), NDGF (Denmark, Norway, Sweden), CC-IN2P3 (France),
KIT/GridKA (Germany), INFN-CNAF (Italy), NL-T1 (Netherlands), PIC (Spain),
ASGC (Taiwan), RAL (UK) and BNL (USA) and in the Tier-2 facilities
worldwide.

%% file: atlas_authlist.tex
\begin{flushleft}
{\Large The ATLAS Collaboration}

\bigskip

G.~Aad$^\textrm{\scriptsize 87}$,
B.~Abbott$^\textrm{\scriptsize 114}$,
J.~Abdallah$^\textrm{\scriptsize 152}$,
O.~Abdinov$^\textrm{\scriptsize 11}$,
B.~Abeloos$^\textrm{\scriptsize 118}$,
R.~Aben$^\textrm{\scriptsize 108}$,
M.~Abolins$^\textrm{\scriptsize 92}$,
O.S.~AbouZeid$^\textrm{\scriptsize 159}$,
H.~Abramowicz$^\textrm{\scriptsize 154}$,
H.~Abreu$^\textrm{\scriptsize 153}$
% ...
\bigskip
\\
% ...
$^{11}$ Institute of Physics, Azerbaijan Academy of Sciences, Baku, Azerbaijan\\
% ...
$^{87}$ CPPM, Aix-Marseille Universit{\'e} and CNRS/IN2P3, Marseille, France\\
% ...
$^{92}$ Department of Physics and Astronomy, Michigan State University, East Lansing MI, United States of America\\
% ...
$^{108}$ Nikhef National Institute for Subatomic Physics and University of Amsterdam, Amsterdam, Netherlands\\
% ...
$^{114}$ Homer L. Dodge Department of Physics and Astronomy, University of Oklahoma, Norman OK, United States of America\\
% ...
$^{118}$ LAL, Univ. Paris-Sud, CNRS/IN2P3, Universit{\'e} Paris-Saclay, Orsay, France\\
% ...
$^{152}$ Institute of Physics, Academia Sinica, Taipei, Taiwan\\
$^{153}$ Department of Physics, Technion: Israel Institute of Technology, Haifa, Israel\\
$^{154}$ Raymond and Beverly Sackler School of Physics and Astronomy, Tel Aviv University, Tel Aviv, Israel\\
% ...
$^{159}$ Department of Physics, University of Toronto, Toronto ON, Canada\\
% ...
\end{flushleft}